\documentclass[twocolumn,showpacs,aps,prc,floatfix,nofootinbib]{revtex4}

\usepackage{graphicx}
\usepackage{dcolumn}
\usepackage{bm}

\begin{document}

\title{Hollow nuclear matter}

\author{Gao-Chan Yong$^{1,2}$}

\affiliation{%
$^1${Institute of Modern Physics, Chinese Academy of Sciences, Lanzhou
730000, China}\\
$^2${State Key Laboratory of Theoretical Physics, Institute of
Theoretical Physics, Chinese Academy of Sciences, Beijing 100190, China}
}

\begin{abstract}
It is generally considered that an atomic nucleus is always compact. Based on the isospin-dependent Boltzmann nuclear transport model, here I show that large block nuclear matter or excited nuclear matter may both be hollow. And the size of inner bubble in these matter is affected by the charge number of nuclear matter.
Existence of hollow nuclear matter may have many implications in nuclear
or atomic physics or astrophysics as well as some practical applications.

\end{abstract}
\maketitle

\section{Introduction}

One generally considers that an atomic nucleus is always compact.
However, it has been theoretically argued that for a nuclear system
the total energy can be decreased if a bubble configuration is created
\cite{Decharg99,Nazarewicz02,Decharg03}.
In fact, the possible exotic bubble
or toroidal configuration of atomic
nuclei has been theoretically discussed for many decades
\cite{Wilson46,Siemens67,wong721,wong722,wong723,wong73,wongrmp,Campi73,Yu00}.
In 1946, Wilson has already studied spherical bubble nuclei \cite{Wilson46}.
In 1967, Siemens and Bethe studied spherical bubble nuclei using a liquid drop model \cite{Siemens67}. And later on, based on a liquid drop model (LDM) with shell correction energy,
Wong studied known $\beta$-stable nuclei and found spherical bubbles \cite{wong73}.
Furthermore, Moretto \emph{et al.} showed that bubbles at finite temperature may be stabilized by the inner vapor pressure \cite{more97}. Borunda and L\'{o}pez, using the
hydrodynamic equations coupled to a simple but realistic equation of state, argued
that the hollow configuration is a direct consequence of the liquid-like behavior of compressed nuclear matter \cite{borun94}.
And transport simulations of nuclear collisions studied by W. Bauer \emph{et al.} indicate the possibility of bubble configuration
in nuclear matter at finite temperature \cite{bauer92}. Recently,
J. Decharg\'{e} \emph{et al.}
performed a self-consistent microscopic Hartree--Fock--Bogoliubov (HFB) calculations using the effective Gogny interaction and found stable bubble solutions
of some specific superheavy nuclei (250 $\lesssim$ Z $\lesssim$ 280, 780 $\lesssim$ A $\lesssim$ 920) \cite{Decharg99,Decharg03}. These results are qualitatively similar to previous studies
based on the liquid drop model (LDM) using Strutinsky shell correction
method \cite{wongrmp,swia66} and phenomenological shell model potentials \cite{Dietrich76,Dietrich98}.

Unlike the probe of bubble formation
in heavy-ion collisions \cite{bauer92}, the hollow configuration of an atomic nucleus is in fact hard to probe \footnote{Based on the
transport model of nucleus-nucleus collisions, the author recently
found that the value of $\pi^{-}/\pi^{+}$ ratio especially its
kinetic energy distribution from the central collision of the
bubble nuclei is evidently larger than that from the normal nuclei
without bubble. The observable $\pi^{-}/\pi^{+}$ ratio thus can be
used to probe nuclear bubble configuration in nucleus.}. Till now the existence of the bubble nucleus is still
in debate. It is thus necessary to re-study whether the
hollow atomic nucleus exists or not by completely different theoretical methods. Based on the isospin-dependent Boltzmann
nuclear transport model (nuclear initialization, single nucleon potential, nucleon-nucleon cross section and Pauli-blocking are all isospin-dependent), here I show that atomic nuclei
with super-large mass number or with excitation
energy may both are hollow.
Besides practical applications, the existence of hollow nuclear matter
may have many implications in quantum many-body theory, nuclear
physics, atomic physics and the configuration of neutron stars, etc.

\section{The Boltzmann nuclear transport model}

To simulate the formation of the hollow configuration of an atomic nucleus, we use
the isospin-dependent Boltzmann nuclear transport model, in which nucleon
co-ordinates in initial nuclei are randomly given in the sphere
with radius $R = 1.2A^{1/3}$ ( $A$ is the mass number of nucleus)
and their momenta are randomly given in the Fermi sea
\cite{bertsch}. The isospin-dependence of nucleon Fermi momentum is considered. We use the Skyrme-type parametrization for the isoscalar mean field, which
reads \cite{bertsch}
\begin{equation}
U(\rho)=A(\rho/\rho_{0})+B(\rho/\rho_{0})^{\sigma}.
\end{equation}
Where $\sigma = 1.3$, A = -232 MeV accounts for attractive part and B =
179 MeV accounts for repulsive part. These choices correspond to an
incompressibility coefficient K = 230 MeV  \cite{todd2005}.
Considering the nucleon-nucleon Short-Range-Correlations (SRC), we let the kinetic
symmetry energy be -6.71 MeV \cite{cli15}. The
symmetry potential becomes \cite{henli14}
\begin{equation}\label{usym}
U^{n/p}_{\rm sym}(\rho,\delta)=38.31(\rho/\rho_0)^{\gamma} \times
[\pm 2\delta+(\gamma-1)\delta^2],
\end{equation}
where $\delta=(\rho_n-\rho_p)/\rho$ is the isospin asymmetry of
nuclear medium and $\gamma$ = 0.3. In the above, we let the value of symmetry energy
at saturation density be 31.6 MeV \cite{esym14,cxu10}. The corresponding symmetry
energy becomes \cite{gli15}
\begin{equation}
E_{sym}=-6.71(\rho/\rho_{0})^{2/3}+38.31(\rho/\rho_{0})^{\gamma}.
\end{equation}

In this model, the in-medium nucleon-nucleon ($NN$) elastic cross
section is factorized by the product of a medium correction
factor and the free baryon-baryon scattering cross section
\cite{liq2006}, i.e.,
\begin{equation}
\sigma_{medium}^{NN,elastic}=(\frac{1}{3}+\frac{2}{3}e^{-u/0.54568})\times
(1\pm0.85\delta)\times\sigma_{free}^{NN,elastic}, \label{elastic}
\end{equation}
where $u = \rho/\rho_{0}$ is the nuclear reduced density and $\rho_{0}$ is nuclear saturation density.
To find the stable state of atomic nucleus
in the simulations, we use 1000 test-particles for each nucleon.

\section{Results and discussions}
\begin{figure}[tbh]
\centering
\includegraphics[width=0.485\textwidth]{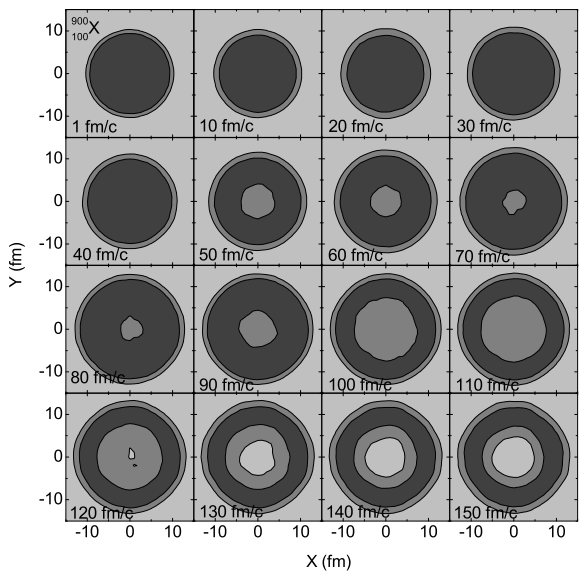}
\caption{(Color online) Time evolution of the contour density
distribution of the hypothetical superheavy atomic nucleus
$^{900}_{100}$X in X - Y plane with the Boltzmann nuclear
transport model. Density becomes larger as color changes from light
to dark. The bubble configuration appears after 50 fm/c in the compact
nucleus $^{900}_{100}$X.} \label{ct900z100}
\end{figure}
\begin{figure}[tbh]
\centering
\includegraphics[width=0.485\textwidth]{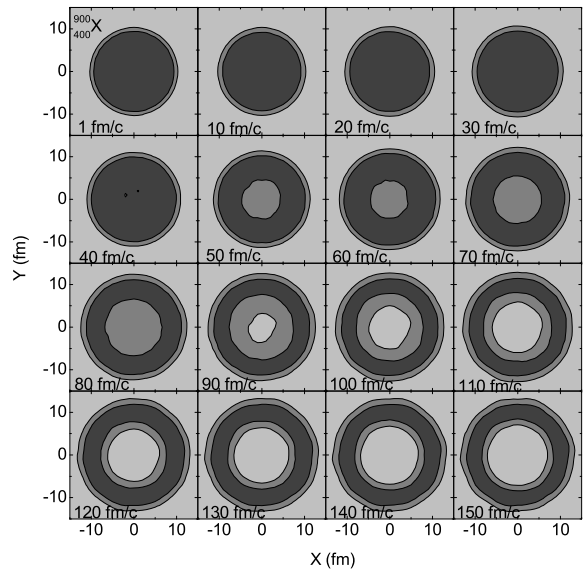}
\caption{(Color online) Time evolution of the contour density
distribution of the hypothetical superheavy atomic nucleus
$^{900}_{400}$X in X - Y plane with the Boltzmann nuclear
transport model. Density becomes larger as color changes from light
to dark. The bubble configuration appears after 40 fm/c in the compact
nucleus $^{900}_{400}$X.} \label{ct900z400}
\end{figure}
With the Gogny effective interaction, J. Decharg\'{e} \emph{et al}.
performed a microscopic self-consistent Hartree--Fock--Bogoliubov
calculations and found stable bubble solutions in the range 250
$\lesssim$ Z $\lesssim$ 280 and 780 $\lesssim$ A $\lesssim$ 920 \cite{Decharg99}.
And these results are qualitatively similar to the results based on the
semi-phenomenological method \cite{Dietrich76,Dietrich98}.
Based on the isospin-dependent Boltzmann nuclear
transport model, I made simulations on the formation of the bubble configuration
of some hypothetical superheavy atomic nuclei. Interestingly, I found the bubble configuration
of hypothetical superheavy atomic nuclei but the charge number dependence is not very evident.
Fig.~\ref{ct900z100} shows the time evolution of the contour density
distribution of the hypothetical superheavy atomic nucleus
$^{900}_{100}$X in X - Y plane with the Boltzmann nuclear
transport model. At the initial time, I fix nucleon co-ordinates
randomly in spheroidal atomic nucleus. It is seen that as time
increases a bubble steadily appears in the compact hypothetical
heavy atomic nucleus. This simulation of atomic nucleus tells us
that such atomic nucleus with large mass number tends to internally
hollow atomic nucleus. Unlike previous discussions in Ref.~%
\cite{Decharg99}, in the simulation, we also see similar internally
hollow atomic nucleus with the same mass number as $^{900}$X$_{100}$ but
different charge number, such as atomic nucleus $^{900}_{400}$X
as shown in Fig.~\ref{ct900z400}. While comparing Fig.~\ref{ct900z400} with
Fig.~\ref{ct900z100}, we find the size of inner bubble depends on the charge number
of the nucleus. This point is somewhat consistent with the study in Ref.~ \cite{Decharg99,Nazarewicz02,Decharg03}.

\begin{figure}[tbh]
\centering
\includegraphics[width=0.485\textwidth]{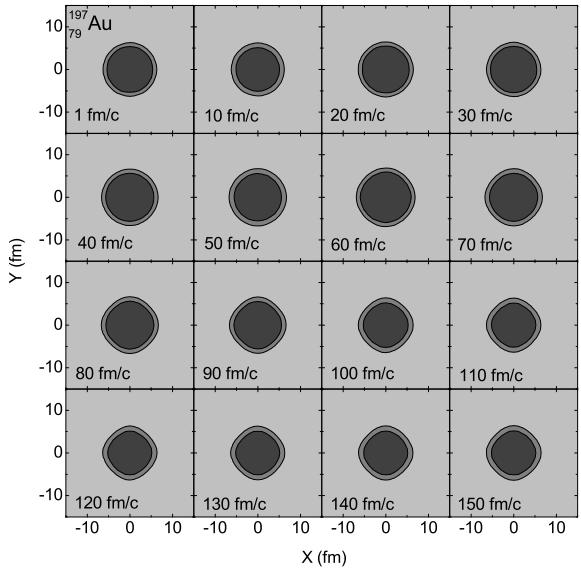}
\caption{(Color online) Time evolution of the contour density
distribution of the relatively light atomic nucleus $^{197}_{79}$Au in
X - Y plane with the Boltzmann nuclear transport model. Density
becomes larger as color changes from light
to dark. The bubble
configuration does not appear in the compact nucleus $^{197}_{79}$Au.}
\label{nct197}
\end{figure}
\begin{figure}[tbh]
\centering
\includegraphics[width=0.485\textwidth]{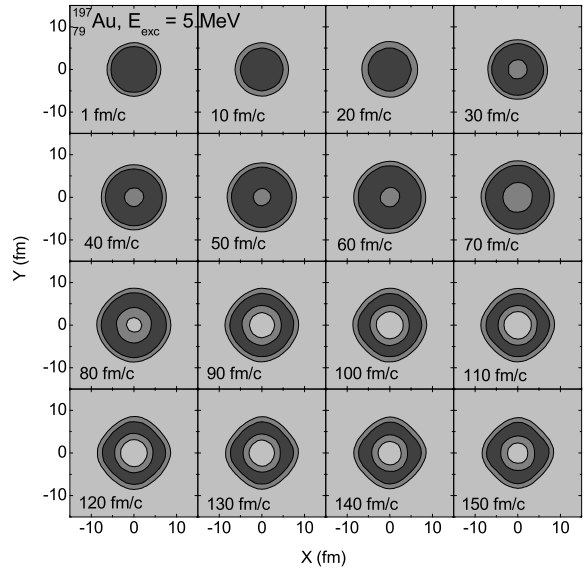}
\caption{(Color online) Time evolution of the contour density
distribution of the relatively light atomic nucleus $^{197}_{79}$Au in
X - Y plane with the Boltzmann nuclear transport model. The giving
excitation energy is average 5 MeV per nucleon. Density becomes
larger as color changes from light
to dark. The bubble configuration
appears after 30 fm/c in the compact nucleus $^{197}_{79}$Au.}
\label{ct197}
\end{figure}
However, for relatively light atomic nucleus such as the stable
existed element $^{197}_{79}$Au, in the evolution of simulation,
we did not see such internally hollow state (as shown in
Fig.~\ref{nct197}) unless giving an excitation energy. Shown in
Fig.~\ref{ct197} is the time evolution of the contour density
distribution of the atomic nucleus $^{197}_{79}$Au in X - Y plane
with the Boltzmann nuclear transport model. The giving average
excitation energy (an average energy per nucleon increased
relative nuclear ground state) is 5 MeV per nucleon. It is seen that
as time increases a bubble steadily appears in the compact light
atomic nucleus $^{197}_{79}$Au. Because the surface tension is
relatively strong for relatively light atomic nucleus, to form
bubble configuration in compact nucleus, one has to give
excitation energy for relatively light atomic nucleus.

Comparing Figs. 1 -- 4, one may conclude that only large block ground-state nuclear matter or
excited nuclear matter are internally hollow. The main reason is that the surface tension is
relatively strong for small block nuclear matter but it becomes weak for large block nuclear matter.

\begin{figure}[tbh]
\centering
\includegraphics[width=0.485\textwidth]{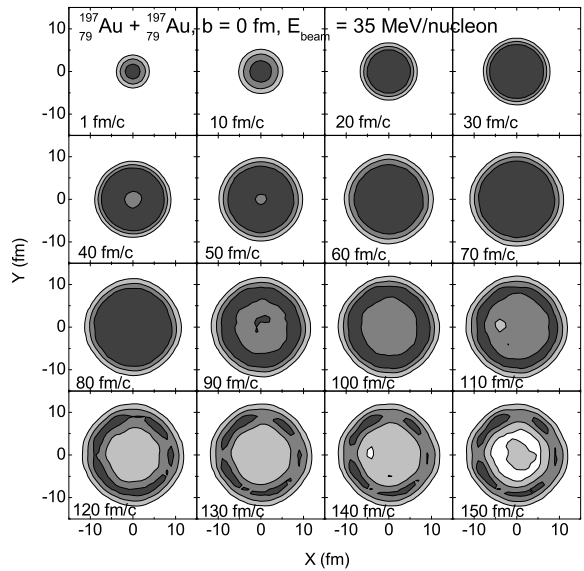}
\caption{(Color online) Time evolution of the contour density
distribution in X - Y plane in the head-on $^{197}_{79}$Au +
$^{197}_{79}$Au collision with the same Boltzmann nuclear
transport model. The incident beam energy (in Z direction) is 35
MeV per nucleon. Density becomes larger as color changes from light
to dark. The bubble configuration of the compressed nuclear matter appears after
90 fm/c.} \label{ct197197}
\end{figure}
The formation of bubble or toroidal nuclear matter was in fact predicted by nuclear transport model at lower incident beam energy \cite{bauer92,liba92,liba93,xu93,zhang14}. It is thus also interesting to see if such internally hollow nuclear matter is
also formed in nucleus-nucleus collisions by our isospin-dependent transport model.
Shown in Fig.~\ref{ct197197} is the
$^{197}_{79}$Au + $^{197}_{79}$Au head-on collision at lower
incident beam energy 35 MeV/nucleon. One can see that internally hollow
nuclear matter is formed in the nucleus-nucleus collisions as time
increases. And inner halo in the bubble of nuclear matter is
also seen \cite{Decharg99,Decharg03}.

\begin{figure}[tbh]
\centering
\includegraphics[width=0.485\textwidth]{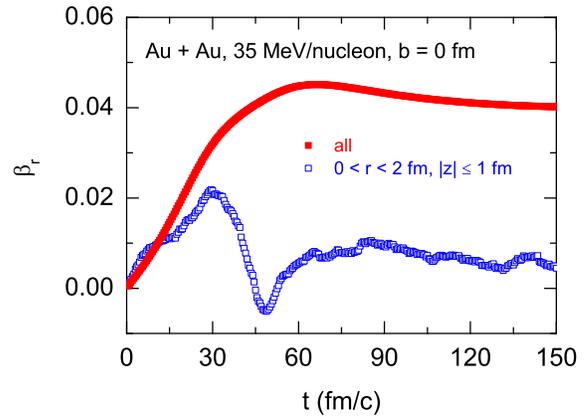}
\caption{(Color online) Time evolution of the radial velocity
$\beta_{r}$ ($= v_{r}/c$) of emitting nucleons in the head-on
$^{197}_{79}$Au + $^{197}_{79}$Au collision with the same
Boltzmann nuclear transport model. The incident beam energy (in Z
direction) is 35 MeV per nucleon. The red solid squares denote
radial velocity of all emitting nucleons while the blue hollow
squares denote radial velocity of inner nucleons with spherical
radius r = 2 fm and $\mid$ Z $\mid$ $\leq$ 1 fm in center of mass
frame.} \label{radialbeta}
\end{figure}
As for how to probe internally hollow nuclear matter formed in heavy-ion collisions,
one can probe the bubble formation by using the radial
velocity of emitting nucleons in
nucleus-nucleus collisions, as shown in
Fig.~\ref{radialbeta}, to deduce whether internally hollow nuclear
matter is formed or not. If the bubble configuration is formed in nucleus-nucleus
collisions, the radial velocity of nucleons in the center of reaction system
should be roughly null due to inversive pressure in the bubble as shown in Fig.~\ref{ct197197} and Fig.~\ref{radialbeta}.

The results presented here are obtained from the isospin-dependent
Boltzmann nuclear transport model, one might question their dependence
upon the particular inputs of the transport model.
In this respect, I made simulations by varying the symmetry energy parameter $\gamma$ from
0.3 to 1.5 (corresponding very soft and stiff symmetry energies). I found the effects
of symmetry energy on the results presented here are very small. While changing the parameters
of the isoscalar mean field, i.e., letting the incompressibility coefficient K = 230 $\rightarrow$ 200 MeV, one sees relatively large changes of the results presented here.
Therefore the incompressibility coefficient used in the transport model needs to be carefully
selected \cite{todd2005}.

Just as the discovery of fullerene cluster in 1985 \cite{Fullerene85}, internally
hollow nuclear matter may have many implications in nuclear
or atomic physics or astrophysics such as the physics of Neutron
stars as well as some practical applications. Existence of internally
hollow atomic nucleus may promote the developments of
quantum many-body theory and nuclear theory.
Increased radius of the hollow atomic nucleus may cause inner
electrons of some atoms to be absorbed easily by inner
nucleons. And the existence of internally hollow large block nuclear matter
may hint the hollow configuration of Neutron stars. Internally hollow
superheavy or excited atomic nuclei may affect the hyperfine configuration
of atomic spectrum, which is widely used in a lot of fields.

\section{Conclusions}

Based on the nuclear transport
model, it is shown that hyperheavy nuclei or general nuclei with excitation energy
may both be internally hollow. And nuclear matter formed in heavy-ion collisions also trends to
inner hollow. One can use the methods of heavy-ion collisions to probe the bubble configuration in nuclear matter. Because the existence of internally
hollow atomic nucleus may promote the developments of
quantum many-body theory and nuclear theory, etc., further studies on the formation as well as its observation signals from nuclear theories and nuclear experiments are urgently needed.

\section*{Acknowledgements}

The work was carried out at National
Supercomputer Center in Tianjin, and the calculations were
performed on TianHe-1A. This work is supported by the National
Natural Science Foundation of China under Grant Nos. 11375239,
11435014.


\begin{thebibliography}{00}

\bibitem{Decharg99}J. Decharg\'{e}, J.F. Berger, K. Dietrich, M.S. Weiss, Phys. Lett. B {\bf 451}, 275 (1999).
\bibitem{Nazarewicz02}W. Nazarewicz et al., Nucl. Phys. A {\bf 701}, 165c (2002).
\bibitem{Decharg03}J. Decharg\'{e}, J.F. Berger, M. Girod, K. Dietrich, Nucl. Phys. A {\bf 716}, 55 (2003).

\bibitem{Wilson46} H. A. Wilson, Phys. Rev. {\bf 69}, 538 (1946).
\bibitem{Siemens67}P. J. Siemens, H. A. Bethe, Phys. Rev. Lett. {\bf 18}, 704 (1967).

\bibitem{wong721}C. Y. Wong, Phys. Lett. B {\bf 41}, 446 (1972).
\bibitem{wong722}C. Y. Wong, Phys. Lett. B {\bf 41}, 451 (1972).
\bibitem{wong723}K. T. R. Davies, C. Y. Wong, S. J. Krieger, Phys. Lett. B {\bf 41}, 455 (1972).
\bibitem{wongrmp}M. Brack, J. Damgaard, A. S. Jensen, H. C. Pauli, V. M.
Strutinsky, and C. Y. Wong, Rev. Mod. Phys. {\bf 44}, 320 (1972).
\bibitem{wong73}C.Y. Wong, Ann. Phys. {\bf 77}, 279 (1973).

\bibitem{Campi73}X. Campi, D.W.L. Sprung, Phys. Lett. B {\bf 46}, 291 (1973).
\bibitem{Yu00}Y. Yu, A. Bulgac, P. Magierski, Phys. Rev. Lett. {\bf 84}, 412 (2000).

\bibitem{more97}L. G. Moretto, K. Tso, and G. J. Wozniak, Phys. Rev. Lett.
{\bf 78}, 824 (1997).

\bibitem{borun94}M. Borunda and J. A. L\'{o}pez, Il Nuovo
Cimento {\bf 107A}, 2773 (1994).

\bibitem{bauer92}W. Bauer, George F. Bertsch, and Hartmut Schulz,
Phys. Rev. Lett. {\bf 69}, 1888 (1992).

\bibitem{swia66}W. D. Myers, W. J. Swiatecki, Nucl. Phys. {\bf 81}, 1 (1966).

\bibitem{Dietrich76}K. Dietrich, K. Pomorski, Nucl. Phys. A {\bf 627}, 175 (1997).
\bibitem{Dietrich98}K. Dietrich, K. Pomorski, Phys. Rev. Lett. {\bf 80}, 37 (1998).

\bibitem{bertsch}G. F. Bertsch, S. Das Gupta, Phys. Rep., {\bf
160}, 189 (1988).

\bibitem{todd2005}B. G. Todd-Rutel, J. Piekarewicz,
Phys. Rev. Lett. {\bf 95}, 122501 (2005).

\bibitem{cli15}B. J. Cai, B. A. Li, Phys. Rev. C {\bf 92}, 011601 (2015).
\bibitem{henli14}O. Hen, B. A. Li, W. J. Guo, L. B. Weinstein, E. Piasetzky, Phys. Rev. C {\bf 91}, 025803 (2015).
\bibitem{esym14}J. Xu, L. W. Chen, B. A. Li, H. R. Ma, Astrophys .J. {\bf 697},1549 (2009).
\bibitem{cxu10}C. Xu, B. A. Li, L. W. Chen, Phys. Rev. C {\bf 82}, 054607 (2010).
\bibitem{gli15}B. A. Li, W. J. Guo, Z. Z. Shi, Phys. Rev. C {\bf 91}, 044601 (2015).
\bibitem{liq2006}Q. F. Li, Z. X. Li, S. Soff, M. Bleicher and H. St\"{o}cker, J. Phys. G {\bf 32}, 407 (2006).

\bibitem{liba92}D. H. E. Gross, Bao-An Li, A. R. DeAngelis, Annalen der Physik, {\bf 504}, 467 (1992).
\bibitem{liba93}Bao-An Li, D. H. E. Gross, Nuclear Physics A {\bf 554}, 257 (1993).
\bibitem{xu93}H. M. Xu, J. B. Natowitz, C. A. Gagliardi, R. E. Tribble, C. Y. Wong, W. G. Lynch, Phys. Rev. C {\bf 48}, 933 (1993).
\bibitem{zhang14} K. Cherevko, L. Bulavin, J. Su, V. Sysoev and F. S. Zhang, Phys. Rev. C {\bf 89}, 014618 (2014).

\bibitem{Fullerene85}H. W. Kroto, J. R. Heath, S. C. Obrien, R. F. Curl, R. E. Smalley, Nature {\bf 318}, 162 (1985).


\end{thebibliography}
\end{document}